\documentclass[12pt,a4paper]{article}

\usepackage[utf8]{inputenc}
\usepackage[T1]{fontenc}
\usepackage{amsmath,amssymb,bm,graphicx}
\usepackage[colorlinks=true,linkcolor=blue,citecolor=red]{hyperref}
\usepackage{geometry}
\usepackage{lineno}
\usepackage[sort&compress,numbers]{natbib}
\usepackage{physics}
\usepackage{braket}
\usepackage{booktabs}
\usepackage{array}
\usepackage{amsfonts}
\usepackage{authblk}

\title{The Equivalence Principle and Kinematical Structure in the ADM Framework}
\author{L. G. Pereira}
\affil{\small \textit{Instituto de Física, Universidade Federal do Rio Grande do Sul (UFRGS) \\
Av. Bento Gonçalves 9500, Caixa Postal 15051 \\
Porto Alegre, RS 91501-970, Brazil \\
\texttt{lgp@if.ufrgs.br}}}
\date{}

\begin{document}


\maketitle

\begin{abstract}
The relation between uniformly accelerated laboratories and laboratories
supported in a gravitational field lies at the conceptual core of the
Equivalence Principle, yet its precise kinematical content beyond strictly
local considerations remains subtle. In this work we develop a unified
metric description of these configurations using the standard
Arnowitt–Deser–Misner (ADM) formulation of General Relativity, which
provides an explicit decomposition of spacetime into spatial hypersurfaces
and their temporal evolution. In this setting the ADM shift vector is
interpreted as a physical quantity encoding the kinematical relation
between spatial slices and their temporal embedding associated with a
chosen foliation. This interpretation allows uniformly accelerated
laboratories and laboratories supported in a gravitational field to be
described within a common structural framework, showing that configurations
experiencing identical proper acceleration share an equivalent local shift
structure. This viewpoint clarifies the apparent asymmetry between the spatial
displacement and energetic cost associated with accelerated motion and
their apparent absence in phenomenological descriptions of observers
supported in a gravitational field. The formulation remains fully equivalent to standard General Relativity at the level of the field equations and constraints while making explicit kinematical features that are usually implicit in its geometric description. Consequences of this interpretation include a relational
account of gravitational time dilation and the emergence of
observer-dependent horizons.

\vspace{1em}
\end{abstract}

\section{Introduction}

The Equivalence Principle stands as one of the most profound conceptual insights in the history of physics. Its roots lie in the early recognition of the universality of free fall, notably in Galileo’s investigations, and in Newton’s formulation of mechanics, where inertial and gravitational mass enter with empirically indistinguishable numerical values. Although this equality functioned operationally within Newtonian theory, its deeper significance remained conceptually unexplained. It was Einstein who elevated this empirical fact to the status of a foundational physical principle, recognising that a laboratory at rest in a uniform gravitational field (in the appropriate local sense) is locally indistinguishable from one undergoing uniform acceleration. This insight—later described by Einstein as ``the happiest thought of my life''\footnote{See, for example, Pais \cite{pais1982} for a historical discussion.}—provided the conceptual bridge from a force-based description of gravity to its geometric interpretation \cite{einstein1907,einstein1911,einstein1916}. Within the geometric formulation of General Relativity, this intuition is encoded in the statement that free-falling bodies follow geodesics determined by the spacetime metric, whereas non-geodesic motion reflects a deviation from the local inertial structure. Over time, distinct formulations—weak, Einstein, and strong—have been articulated, each emphasising different aspects of the deep connection between inertia and gravitation \cite{will2014,poisson2014}. In the geometric viewpoint originally articulated by Einstein \cite{einstein1916}, developed in modern form \cite{misner1973, wald1984}, and empirically validated \cite{will2014}, gravitational effects may be locally eliminated by an appropriate choice of coordinates, with any residual phenomena attributed to spacetime curvature.

 Despite its empirical success and universal acceptance, the Equivalence Principle, in its conventional geometric formulation, exhibits a persistent interpretative tension once one moves beyond strictly local considerations. This tension becomes particularly transparent when comparing uniformly accelerated laboratories with laboratories statically supported in a gravitational field. Although both measure identical proper acceleration (at corresponding worldlines), their global descriptions appear asymmetric. This asymmetry—frequently discussed in the literature as a tension between local and global equivalence \cite{norton1985,janssen2005}—will be referred to in this work as the \emph{distance paradox}: an accelerated laboratory actively accumulates spatial separation relative to inertial observers, whereas a gravitationally supported laboratory appears stationary, despite exhibiting physically analogous effects. A second physical difference between these configurations concerns their energetic character. A uniformly accelerated rocket requires a continuous expenditure of energy to maintain its state of motion, whereas a laboratory supported at the surface of a gravitating body appears, at least in its standard description, not to incur an analogous local energetic cost. We shall refer to this tension as the \emph{energy paradox}. 
Furthermore, analyses of non-inertial frames—even in flat spacetime—demonstrate that several phenomena commonly attributed to gravity, such as horizons, redshifts, and causal restrictions, may arise purely from the kinematical state of the observer \cite{rindler1966,brown2005,unruh1976}. This observation suggests that part of the physical content of the Equivalence Principle may be understood as intrinsically kinematical in character.

Within this context, if the two laboratories are to be regarded as physically equivalent in the sense of the Equivalence Principle, they must be treated on an equal structural footing, that is, within a unified metric description applicable to both situations: the uniformly accelerated rocket and the laboratory supported in a gravitational field. For the rocket, the Rindler metric already provides a purely kinematical description; for the supported laboratory, the Schwarzschild solution is traditionally interpreted, in its standard reading, in strictly geometrical terms. The central challenge addressed in this work is to formulate a single metric framework capable of encompassing both configurations, even though they appear as distinct solutions within the standard formulation of General Relativity. Two broad strategies naturally present themselves. One may attempt to “geometrise” the Rindler metric, attributing its features to spacetime curvature; alternatively, one may “kinematise” the Schwarzschild solution, reinterpreting its gravitational features in explicitly kinematical terms. In the present work, we pursue the second strategy. The first step towards such a unified description is to adopt a framework that distinguishes space and time while preserving the full mathematical structure of General Relativity. The Arnowitt–Deser–Misner (ADM) formulation—originally developed as a Hamiltonian formulation of the theory—fulfils precisely this role \cite{arnowitt1959,arnowitt1962,gourgoulhon2012,alcubierre2008}. It provides a decomposition of spacetime into spatial hypersurfaces, their intrinsic geometry, and their embedding in time.

In our proposal, the ADM formalism is employed without modification and remains fully equivalent to standard General Relativity at the level of field equations and constraints; no new dynamics is introduced. However, the shift vector is interpreted not merely as a coordinate artefact, nor as representing a physical flow of matter, but as a structural field encoding the local inertial baseline associated with a chosen foliation. Within this perspective, the ADM shift vector captures the kinematical relation between space and time that underlies both accelerated and gravitational configurations.

The paper is organised as follows. In Section \ref{sec02}, we introduce the ADM formalism, articulate the shift hypothesis, and demonstrate how this framework preserves the standard relativistic invariants. We then apply this structure to address both the distance and energy paradoxes through a characterisation of non-inertial configurations. In Section \ref{sec03}, we examine the direct physical consequences of this kinematical structure, including a relational derivation of gravitational time dilation, the emergence of observer-dependent horizons, and a kinematical account of the directional character associated with the cumulative structure of temporal and ADM spatial displacements. Finally, Section \ref{conc} presents our concluding remarks on how this unified formulation clarifies the structural role of the Equivalence Principle.

\section[The ADM formalism and the shift hypothesis]
{The ADM formalism and the shift hypothesis%
}
\label{sec02}

The Arnowitt--Deser--Misner (ADM) formalism~\cite{arnowitt1959,arnowitt1962}
provides a decomposition of spacetime into spatial hypersurfaces and their
temporal evolution. Beyond its original Hamiltonian motivation, the ADM
framework is particularly suited for the present investigation because it
allows one to disentangle spatial geometry, temporal structure, and the
kinematical relation between them \footnote{For further details on the Initial-Value Problem and this formalism, see Misner, Thorne, and Wheeler~\cite{misner1973}, Chapter 21.}.

The ADM line element is written as
\begin{equation}
ds^{2}
= -N^{2} dt^{2}
+ h_{ij}\big(dx^{i} + N^{i} dt\big)\big(dx^{j} + N^{j} dt\big),
\label{eq:ADM-basic}
\end{equation}
where $N$ is the lapse function, $N^{i}$ the shift vector, and $h_{ij}$ the
induced spatial metric on constant--time hypersurfaces.

Within our proposed framework, where the intrinsic spatial geometry is taken to be locally flat and the metric must naturally recover the Minkowskian form in the appropriate inertial limit, all nontrivial kinematical effects are encoded exclusively in the shift vector, leading to the following identities for the ADM parameters:
\begin{enumerate}
    \item The spatial geometry is locally flat,
    \begin{equation}
        h_{ij} = \delta_{ij}.
    \end{equation}
    \item The lapse function is fixed at its inertial value,
    \begin{equation}
        N = c,
    \end{equation}
    so that the coordinate time $t$ coincides locally with the proper time of
    observers comoving with the chosen foliation. This choice isolates all
    nontrivial effects in the kinematical relation between space and time.
    \item The shift vector is written as
    \begin{equation}
        N^{i} = v^{i}.
    \end{equation}
\end{enumerate}

Under these assumptions, the ADM line element reduces to
\begin{equation}
\boxed{
ds^{2}
= -\!\left(c^{2} - v^{2}\right) dt^{2}
+ 2\,v_{i}\, dx^{i} dt
+ \delta_{ij}\, dx^{i} dx^{j},
}
\label{eq:ADM-unified}
\end{equation}
where $v^{2}=v_{i}v^{i}$ with respect to the Euclidean spatial metric.

The shift vector $N^i$ plays a central role in the present work. In the formulation developed here, it will not be treated as a single elementary object. Instead, we introduce a decomposition that makes explicit the distinct kinematical contributions associated with the chosen foliation and with the motion of laboratory observers.

Our proposal is to decompose the shift as
\begin{equation}
N^i = v^i_{\mathrm{ADM}} + u^i.
\label{ui}
\end{equation}

The term $v^i_{\mathrm{ADM}}$ represents the shift associated with a
non--inertial foliation, independently of the physical origin of the non--inertial configuration. For an accelerated laboratory, such as a rocket, this term
corresponds to $v^i_{\mathrm{R}}$, while for a laboratory held at rest in a
gravitational field it corresponds to $v^i_{\mathrm{G}}$. In both cases,
$v^i_{\mathrm{ADM}}$ characterises the kinematical structure of the foliation
itself, rather than any material motion of the laboratory.

The second contribution is defined as the projection of the
phenomenological relational velocity onto the ADM foliation; 
this necessarily requires at least two laboratories. Consider therefore two laboratories, $A$ and $B$. All velocities discussed below are relational quantities defined with respect to this pair.

The ADM relational velocity of $B$ as measured by $A$ is defined as
\begin{equation}
u_{BA}^i \equiv \frac{dx_B^i}{d\tau_A},
\end{equation}
where $x_B^i$ denotes the ADM spatial coordinates of $B$, and
$\tau_A$ is the proper time of laboratory $A$. This quantity
describes the displacement of $B$ along the ADM spatial grid
as parametrised by the clock of $A$.

Independently of the ADM foliation, we define the phenomenological
relational velocity of $B$ with respect to $A$ as
\begin{equation}
w_{BA}^i \equiv \frac{dr_B^i}{d\tau_A},
\end{equation}
where $r_B^i$ denotes the operational spatial distance from the
center of mass that sources the shift field. The coordinate $r_B^i$
is thus defined relative to the physical mass distribution and
is independent of the choice of ADM spatial coordinates.

Applying the chain rule yields the general relation
\begin{equation}
w_{BA}^i
=
\frac{dr_B^i}{dx_B^j}
\frac{dx_B^j}{dt_A}
\frac{dt_A}{d\tau_A}
=
\left( \frac{dr_B^i}{dx_B^j} \right)
\left( \frac{dt_A}{d\tau_A} \right)
u_{BA}^j.
\end{equation}

This expression is purely kinematical and holds for arbitrary
configurations of $A$ and $B$. 

In the limit where both laboratories are inertial and at relative rest in Minkowski spacetime, their spatial separation is constant. Hence $dr^i = 0$ and $dx^i = 0$, implying $w^i = 0$ and $u^i = 0$. Consequently, the unified metric (Eq.~\ref{eq:ADM-unified}) reduces to
\begin{equation}
ds^2 = -c^2 dt^2 + \delta_{ij} dx^i dx^j,
\end{equation}
which is precisely the Minkowski line element.\footnote{The quantity $\frac{dr}{dx}$ is determined by the spatial mapping associated with the chosen foliation. It represents a geometric conversion factor between the radial coordinate $r$ and the ADM spatial coordinate $x$, defined pointwise on each hypersurface (analogous to a Jacobian factor). It is structural and does not encode dynamical information about motion along a worldline. For instance,  in the foliation of an inertial laboratory ($\mathrm{Lab}_\mathrm{I}$), one has $x=r$, so that $dr/dx = 1$ identically, independently of the state of motion.}

\subsection{Preservation of relativistic invariants}

Two fundamental invariants define the kinematical structure of General
Relativity: the existence of an invariant spacetime interval,
\begin{equation}
ds^{2} = g_{\mu\nu} dx^{\mu} dx^{\nu},
\end{equation}
and the existence of an invariant causal speed $c$, characterizing null
directions of the metric.

Within the present framework, the spacetime geometry is described by the
unified ADM line element \eqref{eq:ADM-unified}.
The invariance of the interval follows trivially from its definition as a
scalar under coordinate transformations. Nevertheless, it is instructive to
verify explicitly that the presence of the shift does not modify the causal
structure encoded in the metric.

For null trajectories, $ds^{2}=0$, Eq.~\eqref{eq:ADM-unified} yields
\begin{equation}
\delta_{ij}\frac{dx^{i}}{dt}\frac{dx^{j}}{dt}
+ 2 v_{i}\frac{dx^{i}}{dt}
- \left(c^{2}-v^{2}\right)
=0.
\end{equation}
Rearranging terms,
\begin{equation}
\left(\frac{dx^{i}}{dt}+v^{i}\right)
\left(\frac{dx_{i}}{dt}+v_{i}\right)
= c^{2},
\end{equation}
or equivalently,
\begin{equation}
\frac{dx^{i}}{dt}
= -v^{i} \pm c\, n^{i},
\label{eq:null-velocity}
\end{equation}
where $n^{i}$ is a unit spatial vector satisfying
$\delta_{ij}n^{i}n^{j}=1$.

Equation~\eqref{eq:null-velocity} shows that while the coordinate velocity is anisotropic due to the shift, the causal wavefront propagates at $c$ relative to the local kinematical structure. The shift field modifies the coordinate relation between spatial and temporal intervals but preserves the invariant null character of the interval.

Therefore, the metric \eqref{eq:ADM-unified} preserves both fundamental
relativistic invariants: the spacetime interval $ds^{2}$ and the universal
speed $c$. The presence of the mixed space--time term does not signal a
modification of relativistic physics, but a reorganization of the splitting
between space and time induced by a non--inertial foliation.

\subsubsection{Vacuum solutions and the shift profile}

We now determine the specific form of the shift field $v(x)$ required by Einstein's equations. For a stationary metric with planar symmetry (one--dimensional spatial dependence), the vacuum Einstein equations ($R_{\mu\nu}=0$) reduce to a simple constraint on the kinematical structure of the foliation, expressed through the spatial variation of the squared ADM shift:
\begin{equation}
\frac{d^{2}}{dx^{2}}\left(v^{2}(x)\right) = 0.
\end{equation}

The general solution is linear, $v^{2}(x) = 2a(\lambda-x)$. This profile corresponds exactly to the Rindler metric expressed in Painlevé-Gullstrand coordinates, describing a frame with uniform proper acceleration $a$.

Conversely, imposing spherical symmetry, as appropriate for a pointlike mass
distribution, Einstein’s vacuum equations reduce to the constraint
\begin{equation}
\frac{d}{dr}\!\left(r\, v^{2}(r)\right)=0 .
\end{equation}
Its unique solution is
\begin{equation}
v(r)=\sqrt{\frac{2GM}{r}},
\end{equation}
which corresponds to the Painlevé--Gullstrand \cite{painleve1921,gullstrand1922} representation of the
Schwarzschild geometry. Thus, the shift profiles for both the accelerated laboratory (linear) and the gravitational field (inverse square) arise as the unique vacuum solutions for their respective symmetries within the ADM framework.

\subsubsection{The distance paradox}

Consider two laboratories undergoing the same constant proper acceleration
$a$: an accelerated laboratory (Lab$_\mathrm{R}$) and a laboratory held at rest
in a gravitational field (Lab$_\mathrm{G}$). In both cases, the laboratory
worldline can be represented as a stationary observer in the chosen ADM
foliation, satisfying
\begin{equation}
dx^i = 0 .
\end{equation}

Under the assumptions adopted, the spacetime metric is given
by Eq.~\eqref{eq:ADM-unified}. For stationary configurations depending on a single spatial coordinate $x$,
Einstein's vacuum equations impose a direct constraint on the shift profile.
Independently of any assumption about the laboratory trajectory, the Einstein
tensor yields the kinematical relation
\begin{equation}
a = -\frac{1}{2}\frac{d}{dx} v^2(x),
\end{equation}
a result well known in stationary slicings of flat and Schwarzschild spacetimes
\cite{rindler1966,martel2001,hamilton2008,natario2009}. Evaluating the shift along the laboratory worldline
$x=x_0$ then yields

\begin{equation}
v^2_{\mathrm{Lab}} = 2a(\lambda - x_0),
\label{eq:v-lab}
\end{equation}
which is constant, as required for stationary motion.

The parameter $x_0$ labels the spatial coordinate of the laboratory within the
foliation. Since both laboratories are defined as stationary observers and physical predictions depend only on coordinate intervals, this alignment does not imply a global identification of distinct foliations.

Under this local identification,
\begin{equation}
x_{0\mathrm{R}} = x_{0\mathrm{G}} \equiv x_0.
\end{equation}

For stationary foliations associated with uniform proper acceleration, the
Einstein equations admit a unique shift profile up to trivial translations,
so that $\lambda$ is completely fixed once $a$ is specified.
Therefore, imposing
\begin{equation}
a_{\mathrm{R}} = a_{\mathrm{G}},
\end{equation}
implies
\begin{equation}
\lambda_{\mathrm{R}} = \lambda_{\mathrm{G}} \equiv \lambda .
\label{eq:vprofile}
\end{equation}

Under these conditions, Eqs.~\eqref{eq:v-lab} and~\eqref{eq:vprofile} immediately lead to
\begin{equation}
v^i_{\mathrm{R}} = v^i_{\mathrm{G}} .
\end{equation}

Since the shift fields coincide pointwise along the laboratory worldlines,
$v_{\mathrm R}(t)=v_{\mathrm G}(t)$, the associated kinematical displacements
satisfy
\begin{equation}
\Delta x_R = \int_{0}^{\Delta t} v_R(t)\,dt = \int_{0}^{\Delta t} v_G(t)\,dt=\Delta x_G.
\end{equation}

This expression refers to a displacement induced by the kinematical structure of
the spacetime description and should not be interpreted as a material
trajectory.

This result establishes the distance paradox: laboratories subject to
physically distinct mechanisms—acceleration by a rocket or support in a
gravitational field—are described by identical kinematical shift structures
when they share the same proper acceleration. The paradox reflects not a
degeneracy of physical situations, but the fact that the shift encodes the
kinematical properties of the spacetime foliation rather than material motion
or force generation.

\subsubsection{The energy paradox}

The energetic characterisation of non-geodesic laboratories follows directly
from the preservation of the relativistic invariants $c=\mathrm{const}$ and
$ds^{2}=\mathrm{const}$. For stationary spacetimes, the existence of a timelike
Killing vector $\xi^{\mu}=(\partial_t)^{\mu}$ allows one to define a conserved
energy per unit mass along any worldline\footnote{
Conserved energies associated with timelike Killing vectors and their
evaluation for stationary observers in metrics with nonvanishing shift are
standard results in General Relativity. For derivations closely related
to the Painlev\'e--Gullstrand/Martel–Poisson form of the Schwarzschild
geometry, see, e.g., Refs.~\cite{martel2001,hamilton2008,wald1984,misner1973,poisson2014}.
},
\begin{equation}
E = -u_{\mu}\xi^{\mu} = -g_{t\mu}u^{\mu}.
\label{eq:energy-def}
\end{equation}

For the ADM metric
\begin{equation}
ds^{2}
= -(c^{2}-v^{2})dt^{2}
+2v_i dx^i dt
+\delta_{ij}dx^i dx^j,
\end{equation}
consider stationary observers in the chosen foliation, defined operationally by
\begin{equation}
dx^i = 0 .
\end{equation}
Their four--velocity reads
\begin{equation}
u^{\mu}=\left(\frac{dt}{d\tau},0,0,0\right).
\end{equation}

Along such worldlines, the invariant interval reduces to
\begin{equation}
ds^{2}=-c^{2}d\tau^{2}=-(c^{2}-v^{2})dt^{2},
\end{equation}
from which one obtains
\begin{equation}
\frac{dt}{d\tau}=\frac{c}{\sqrt{c^{2}-v^{2}}}.
\label{eq:dtaudt}
\end{equation}

Substituting Eq.~\eqref{eq:dtaudt} into Eq.~\eqref{eq:energy-def} yields
\begin{equation}
E
=
(c^{2}-v^{2})\frac{c}{\sqrt{c^{2}-v^{2}}}
=
c\sqrt{c^{2}-v^{2}}
=
c^{2}\sqrt{1-\frac{v^{2}}{c^{2}}}.
\label{eq:energy-v}
\end{equation}

Under the assumption of stationarity, this conserved Killing energy per unit mass depends exclusively on the shift magnitude evaluated along the worldline. It should not be confused with special-relativistic kinetic energy; rather, it arises from the projection $E=-u_\mu\xi^\mu$ and is completely determined by the kinematical structure of the foliation.

From the distance paradox discussed previously, laboratories undergoing the
same proper acceleration satisfy
\begin{equation}
v_{\mathrm{R}} = v_{\mathrm{G}},
\end{equation}
which implies that both laboratories possess the same energetic content per unit mass associated with the acceleration field.

The apparent energy paradox arises from the different physical mechanisms that
maintain the non--geodesic state of each laboratory. In the accelerated case
(Lab$_\mathrm{R}$), the departure from inertial motion is sustained by the
rocket engine. In the gravitational case (Lab$_\mathrm{G}$), no local work is
performed on the laboratory. Nevertheless, both situations are described by
the same conserved energy because both share the same shift structure.

The key element underlying this equivalence is the mixed space--time term
$2v_i dx^i dt$ in the metric. This term enforces a kinematical redistribution
between spatial and temporal intervals required to preserve $ds^{2}$ in a
non--inertial foliation. The associated energy should therefore be interpreted
as the asymmetric cost of converting spatial separation into temporal
separation, rather than as mechanical work performed by forces.

The energy paradox is therefore understood at the kinematical level: within the ADM framework, accelerated and gravitational laboratories are described by the same conserved Hamiltonian structure. The energetic cost associated with acceleration or support does not reflect a violation of conservation, but rather the kinematical conversion between spatial and temporal intervals encoded in the shift, while the fundamental relativistic invariants remain preserved.

\subsubsection{Inertial configurations and the ADM shift} 

The previous considerations show that all non--geodesic effects in the present
framework are encoded in the ADM shift field $v^i_{\mathrm{ADM}}$, which
quantifies the kinematical mixing between spatial and temporal directions
induced by a non--inertial foliation. A nonvanishing $v^i_{\mathrm{ADM}}$
reflects a local conversion between spatial and temporal intervals associated with the chosen foliation. For observers whose worldlines are not geodesic with respect to this structure, maintaining such trajectories requires an energetic cost. It is important to distinguish this contribution from purely relational or
phenomenological velocities, which may give rise to additional shift terms
without any intrinsic physical content. Only the ADM component
$v^i_{\mathrm{ADM}}$ encodes the departure from geodesic embedding of the
spatial hypersurfaces.

After these considerations, we adopt the following working assumption: a locally inertial configuration corresponds to a foliation for which there is no net kinematical bias between space and time at the observer's position $x_0$. This is expressed by the pointwise vanishing of the resultant shift vector:
\begin{equation}
v^i_{\mathrm{ADM}}(x_0) = 0 .
\end{equation}

This condition ensures that no cumulative conversion between spatial and temporal intervals takes place at the observer's location, and that no asymmetric energetic cost is required to maintain the laboratory worldline.

In this sense, a locally inertial frame is characterised by the pointwise vanishing of the net ADM shift field. Minkowski spacetime then appears as the simplest global realization of this condition, corresponding to a foliation with trivial kinematical structure and an everywhere-vanishing ADM shift.

\section{Some consequences of the ADM kinematical structure}
\label{sec03}

In this section, we discuss direct and physically observable consequences
of the ADM kinematical structure introduced in the previous section. No new
dynamical assumptions are introduced. All results follow solely from the
existence of a nontrivial shift field and from the preservation of the
relativistic invariants $c=\mathrm{const}$ and $ds^{2}=\mathrm{const}$.

\subsection{Relational time and inertial laboratories}
\label{reltime}
We now compare the proper times $\tau_{\mathrm{I}}$ and $\tau_{\mathrm{L}}$ of two distinct laboratories. Lab$_\mathrm{I}$ is located in an asymptotic region far from any massive sources, while Lab$_\mathrm{L}$ is in radial free fall within the gravitational field generated by a mass $M$. 

To establish a coherent global temporal reference, we adopt the central mass $M$, which acts as the source of the shift field, as the absolute anchor for the global foliation. The kinematical states of both laboratories are then evaluated relationally to this mass.

Both laboratories are locally inertial. Operationally, this is defined by the pointwise vanishing of their intrinsic ADM velocities ($v^i_{\mathrm{ADM}} = 0$). This ensures neither laboratory experiences proper acceleration. Consequently, the total shift $N^i = v^i_{\mathrm{ADM}} + u^i$ evaluated at their respective locations reduces strictly to their relational components, $N^i = u^i$.

Because Lab$_\mathrm{I}$ is defined to be at rest relative to the central mass $M$ in the asymptotic region, its relational velocity is identically zero ($u_{\mathrm{I}} = 0$). In contrast, the free-falling Lab$_\mathrm{L}$ moves relative to $M$, acquiring a nonvanishing relational velocity ($u_{\mathrm{L}} \neq 0$). Since its effective total shift is purely relational ($N^i = u^i$), the unified ADM metric evaluated for Lab$_\mathrm{L}$ takes the exact same structural form as the metric derived in the previous section for an observer with purely intrinsic velocity ($v \neq 0, u = 0$), with $v^i$ simply replaced by $u^i$:
\begin{equation}
ds^2 = -(c^2 - u^2) dt^2 + 2 u_i dx^i dt + \delta_{ij} dx^i dx^j .
\end{equation}
Because the metrics are mathematically isomorphic, Einstein's equations yield the same vacuum solution. For a stationary, spherically symmetric exteriour, the radial constraint equation $\frac{d}{dr}(r u^2) = 0$ strictly determines the magnitude of this relational velocity:
\begin{equation}
u_{\mathrm{L}}(r) \equiv u(r) = \sqrt{\frac{2GM}{r}} .
\label{eq:u_r_falling}
\end{equation}

To quantify the temporal asymmetry, we compare the proper times of both laboratories directly through their invariant intervals within the global foliation. Since both are locally inertial ($v^i_{\mathrm{ADM}} = dx^i/dt = 0$), their temporal evolution is evaluated at constant spatial coordinates within their local comoving frames.

For the reference laboratory Lab$_{\mathrm{I}}$, where the relational velocity vanishes ($u_{\mathrm{I}} = 0$), the metric reduces trivially to:
\begin{equation}
-c^2 d\tau_{\mathrm{I}}^2 = -c^2 dt^2 \quad \implies \quad \frac{d\tau_{\mathrm{I}}}{dt} = 1 .
\label{i-t}
\end{equation}

For the free-falling laboratory Lab$_{\mathrm{L}}$, located at a radial distance $r$ with a nonvanishing relational velocity $u_{\mathrm{L}} = u(r)$, the temporal contribution from the shift modifies the invariant interval:
\begin{equation}
-c^2 d\tau_{\mathrm{L}}^2 = -\left(c^2 - u^2(r)\right) dt^2 \quad \implies \quad \frac{d\tau_{\mathrm{L}}}{dt} = \sqrt{\frac{c^2 - u^2(r)}{c^2}} .
\end{equation}

Taking the direct ratio of these proper rates eliminates the global coordinate time $t$, providing the strict observational relation between the two inertial laboratories:
\begin{equation}
\frac{d\tau_{\mathrm{I}}}{d\tau_{\mathrm{L}}} = \frac{1}{\sqrt{1 - \frac{u^2(r)}{c^2}}} .
\label{l-t}
\end{equation}
Substituting the structural requirement $u(r) = \sqrt{2GM/r}$ from Eq.~\eqref{eq:u_r_falling}, we obtain:
\begin{equation}
\frac{d\tau_{\mathrm{I}}}{d\tau_{\mathrm{L}}} = \left( 1 - \frac{2GM}{r c^2} \right)^{-1/2} > 1 .
\end{equation}

This mathematical result demonstrates a strict non-reciprocity. Although both laboratories are locally inertial ($v^i_{\mathrm{ADM}}=0$) and measure a symmetric relative phenomenological velocity between them ($w_{\mathrm{I|L}} = -w_{\mathrm{L|I}}$), their temporal dilation is not reciprocal. An observer in the free-falling Lab$_{\mathrm{L}}$ concludes that the clock in Lab$_{\mathrm{I}}$ runs faster, while Lab$_{\mathrm{I}}$ observes the clock in Lab$_{\mathrm{L}}$ running slower. This asymmetry confirms that time dilation is governed not by symmetric relative motion, but by the absolute relational state ($u$) of each laboratory with respect to the mass $M$ that anchors the global shift structure.

\subsection{Relative horizons and relational energy}

Building on the kinematical structure anchored to the central mass $M$, we now examine the concept of event horizons from a strictly relational perspective. We consider an inertial, free-falling laboratory, denoted Lab$_{\mathrm{obs}}$, located at a finite radial position $r_{\mathrm{obs}}$. Its kinematical state is characterised by the nonvanishing relational velocity $u_{\mathrm{obs}} \equiv u(r_{\mathrm{obs}})$ relative to $M$.

A relative event horizon for this observer is defined as the radial position $r_{\mathrm{H,obs}}$ from which an outgoing light signal cannot propagate outward relative to Lab$_{\mathrm{obs}}$. In the global foliation, radial null propagation is governed by the invariant local signal speed $c$, which is advected by the underlying shift field. Consequently, the boundary of causal communication between a deeper free-falling emission point and Lab$_{\mathrm{obs}}$ occurs when the relative relational velocity between these two locations exactly equals the speed of light.

This relative horizon condition is expressed directly as:
\begin{equation}
u(r_{\mathrm{H,obs}}) - u_{\mathrm{obs}} = c \quad \implies \quad u(r_{\mathrm{H,obs}}) = c + u_{\mathrm{obs}} .
\end{equation}
Substituting the established radial profile $u(r) = \sqrt{2GM/r}$, we obtain:
\begin{equation}
\sqrt{\frac{2GM}{r_{\mathrm{H,obs}}}} = c + \sqrt{\frac{2GM}{r_{\mathrm{obs}}}} ,
\end{equation}
which yields the relative horizon radius:
\begin{equation}
r_{\mathrm{H,obs}} = \frac{2GM}{\left(c + \sqrt{\frac{2GM}{r_{\mathrm{obs}}}}\right)^2} .
\label{eq:relative_horizon}
\end{equation}

Equation~\eqref{eq:relative_horizon} demonstrates that the event horizon is not an absolute geometric surface, but a relational construct depending on the kinematical state of the observer relative to the mass $M$. For any free-falling laboratory at a finite $r_{\mathrm{obs}}$, the associated horizon lies strictly inside the standard absolute horizon defined for an asymptotic observer ($r \rightarrow \infty$, where $u_{\mathrm{obs}} = 0$).

It is important to emphasise that this does not imply information can freely escape a given event horizon. For Lab$_{\mathrm{obs}}$, the relative horizon $r_{\mathrm{H,obs}}$ remains a strict one-way causal boundary: signals originating inside this radius cannot reach this specific observer. However, because the horizon is observer-dependent, information may propagate outward across successive neighborhoods, provided it is relayed between free-falling laboratories with different relational velocities. In this sense, information is not blocked by a single fixed global surface, but propagates through a chain of overlapping causal domains.

The apparent loss of information arises only when the specific horizon associated with an asymptotic observer ($u=0$) is artificially elevated to an absolute status for all observers. Within the present framework, information preservation is therefore governed not by a rigid global geometry, but by the energetic and kinematical relations required to sustain signal propagation between shifting causal domains.

\subsection{Considerations on temporal and spatial accumulation}

We now examine how the kinematical structure introduced above bears on the nature of time, focusing on the asymmetry that emerges from the presence of a nonvanishing shift field.

\subsubsection{Non-reciprocal proper-time accumulation}

Recall from Subsec.~\ref{reltime} the comparison between the asymptotic inertial laboratory Lab\(_\mathrm{I}\) (Eq. \ref{i-t}) and the freely falling laboratory Lab\(_\mathrm{L}\) (Eq. \ref{l-t}). Over a coordinate interval \(\Delta t = t_2 - t_1\), the accumulated difference in proper time is
\begin{equation}
\Delta \tau \equiv \int_{t_1}^{t_2} \left( \frac{d\tau_\mathrm{I}}{dt} - \frac{d\tau_\mathrm{L}}{dt} \right) dt 
= \int_{t_1}^{t_2} \left( 1 - \sqrt{1 - \frac{u^2(r(t))}{c^2}} \right) dt > 0 .
\end{equation}
This quantity is strictly positive and grows monotonically as Lab\(_\mathrm{L}\) falls inward (where \(u(r)\) increases). It measures the cumulative effect of the shift: proper time is systematically ``lost'' relative to the asymptotic clock, and this loss cannot be recovered without changing the direction of the flow.

\subsubsection{Closed radial cycle and non-closure in the structural variable \texorpdfstring{$x$}{x}}

Here we analyse a radial cycle that is closed in the phenomenological coordinate $r$, in two parts (where $N^i $ is positive in the direction of mass M).

\paragraph{Free fall (descent).}

In free fall the laboratory follows the shift. In the ADM formulations, we have
\begin{equation}
u_1 = v_G.
\end{equation}

The structural displacement is therefore
\begin{equation}
\Delta x_1
=
\int_{t_a}^{t_b} v_G\, dt > 0.
\end{equation}

In other words, it is a positive quantity.

\paragraph{Powered ascent.}

During ascent, the proper acceleration of the rocket kinematically induces an additional structural shift $v_R$ directed towards its floor (in this case, aligned with the same direction as $v_G$).
The total structural shift is
\begin{equation}
u_2 = v_G + v_R.
\end{equation}

The structural displacement becomes
\begin{equation}
\Delta x_2
=
\int_{t_b}^{t_c} (v_G + v_R)\, dt > 0.
\end{equation}

\paragraph{Total structural displacement.}

The total structural displacement over the complete radial cycle is
\begin{equation}
\Delta x_{\mathrm{total}} = \Delta x_1 + \Delta x_2 > 0.
\end{equation}

Although the trajectory is closed in the phenomenological coordinate $r$,
it is not necessarily closed in the structural variable $x$.
The non-closure is therefore a structural consequence of the shift dynamics,
not a coordinate artefact.

\section{Conclusions}
\label{conc}

The objective of this work was to establish a metric framework in which
both situations associated with the Equivalence Principle can be
described within the same spacetime structure. In particular, we address
an apparent interpretative asymmetry between two configurations: an
accelerated rocket, where a spatial displacement is accumulated and an
energetic cost is required to sustain the motion, and an observer
maintained at rest on the surface of a massive body, where neither of
these effects is phenomenologically observed. This apparent asymmetry
has frequently been interpreted as evidence that their equivalence is
merely local \cite{einstein1907,einstein1916,norton1985,brown2005}.
We show that this asymmetry disappears once spacetime is described
within a kinematical framework that makes the foliation structure
explicit.

Starting from the ADM decomposition and a reinterpretation of the shift vector as a composite field defining the local inertial baseline of the foliation, we obtained a unified description of non-inertial configurations. Both $Lab_R$ and $Lab_G$, despite their distinct global settings, are characterised by the same kinematical condition, namely a non-vanishing relative velocity with respect to the ADM shift field ($v_{\mathrm{ADM}} \neq 0$). The spatial displacement of the rocket and the static support of the laboratory are thus recognised as different manifestations of a single underlying mechanism: the deviation from the natural evolution of the spatial foliation. Within this perspective, the proper acceleration measured in either laboratory—whether generated by propulsion or by mechanical support—can be interpreted as the energetic cost associated with preventing the kinematical conversion between spatial and temporal intervals induced by the shift. The inertial state is therefore naturally characterised by the condition $v^i_{\mathrm{ADM}} = 0$, identifying the freely falling laboratory ($Lab_L$) as the configuration in kinematical rest with respect to the foliation, while supported observers become structurally equivalent to accelerated ones.
On the other hand, when two inertial laboratories are compared
(that is, configurations satisfying $v_{\mathrm{ADM}}=0$),
distinct global structures may still arise. A laboratory in free fall
evolves in position with respect to the mass distribution that
generates the shift field, whereas a laboratory located asymptotically
far from the source remains stationary relative to it. Nevertheless,
both configurations correspond to inertial frames in the sense that
locally the spacetime structure reduces to its Minkowskian limit.
The difference appears only at the relational level: their proper
clocks are not necessarily synchronised, reflecting the distinct
global embedding of their worldlines within the spacetime foliation.

This viewpoint follows directly from Einstein’s original insight
\cite{einstein1907,pais1982}, obtaining the equivalence between inertial
and gravitational phenomena from a statement about the organisation of
spacetime foliation. Since the resistance experienced in non-inertial
situations arises from the same kinematical coupling to the shift, the
traditional equality between inertial and gravitational mass
($m_i = m_g$) no longer appears as an independent postulate requiring
separate justification. Rather than expressing the equivalence of two
fundamentally distinct physical charges, this relation can be understood
as a structural feature of spacetime itself: there exists a single mass
parameter whose dual phenomenological manifestation reflects the way
matter couples to the foliation structure. Conversely, both an
asymptotic laboratory in flat spacetime and a freely falling one in a
gravitational field are characterised by the same structural condition:
the absence of intrinsic kinematical shift ($v^i_{\mathrm{ADM}} = 0$).
In this sense, inertial behaviour is not an intrinsic property of mass,
but a manifestation of being at rest with respect to the spatial
structure defined by the foliation. Although such laboratories may
describe one another relationally as accelerated or desynchronised,
these effects do not signal distinct forms of mass; rather, they arise
from their relative positioning within the same underlying spacetime
organisation. This structural identification of inertial and
gravitational behaviour was already implicit in Einstein’s geometric
formulation; here, however, it is recovered explicitly as a kinematical
property of the foliation within the ADM framework.

The same ADM kinematical framework provides a transparent interpretation of horizon formation. Horizons emerge as relational boundaries defined by the condition that the relative shift velocity equals the signal speed $c$, ensuring causal consistency without the need to introduce absolute geometric barriers. This observer-dependent character of horizon structures parallels well-known analyses of Rindler and black hole horizons \cite{rindler1966,wald1984,unruh1976,rovelli2004}. Furthermore, the kinematical framework reveals an intrinsic directional
asymmetry embedded in the gravitational shift. As demonstrated, a laboratory
can complete a closed phenomenological trajectory ($\oint dr = 0$) only by
expending energy to counter the continuous advection of the spatial mesh.
Consequently, it fails to return to its original structural coordinate
($\Delta x > 0$). This spatial non-closure shows that the coordinate $x$
encodes the accumulated kinematical history associated with navigating the
foliation. Coupled with the monotonic accumulation of temporal desynchronization between inertial observers, this structural drift provides a purely
kinematical mechanism capable of generating temporal asymmetry in
gravitational contexts, without invoking thermodynamic or entropic
assumptions.

Importantly, no modification of General Relativity is proposed. By working within the Painlevé--Gullstrand slicing, the metric remains an exact solution of Einstein’s field equations and reproduces all standard observational predictions. The contribution of this work is therefore interpretative rather than dynamical: by treating the ADM shift as the structural carrier of the inertial flow, the Equivalence Principle acquires a unified and operationally coherent formulation within the existing theory.

\section*{Declarations}

\textbf{Funding:} The author declares that no funds, grants, or other support were received during the preparation of this manuscript.

\textbf{Conflicts of interest:} The author has no relevant financial or non-financial interests to disclose.

\textbf{Data availability:} Data sharing is not applicable to this article as no datasets were generated or analysed during the current study.

\end{document}